\documentclass{iaus}
\usepackage{graphicx}

\title[S265.~Accurate Fundamental Stellar Parameters] %% give here short title %%
{Accurate Fundamental Stellar Parameters} 
\author[Hans Bruntt] %% \& Susanne H{\"o}fner]   %% give here short author list %%
{Hans Bruntt$^{1,2}$}
\affiliation{
$^{1}$LESIA, Observatoire de Paris-Meudon, 92195, France \\[\affilskip]
$^{2}$Sydney Institute for Astronomy, School of Physics, The University of Sydney, NSW, Australia
\\email: {\tt bruntt@phys.au.dk}}

\pubyear{2009}
\volume{265}  %% insert here IAU Symposium No.
\pagerange{xxx--yyy}
\jname{Chemical Abundances in the Universe: Connecting First Stars to Planets}
\editors{K. Cunha, M. Spite \& B. Barbuy, eds.}
\begin{document}

\maketitle

\begin{abstract}
We combine results from interferometry, asteroseismology
and spectroscopic analyses to determine accurate
fundamental parameters (mass, radius and effective temperature) 
of 10 bright solar-type stars covering the H-R
diagram from spectral type F5 to K1. 
Using ``direct'' techniques that are only weakly model-dependent we
determine the mass, radius and effective temperature. 
We demonstrate that model-dependent or ``indirect'' methods 
can be reliably used even for relatively faint single stars for
which direct methods are not applicable. This is important for
the characterization of the targets of the CoRoT and {\em Kepler} space missions.

% spectroscopic analysis that rely in atmospheric models)
%This is only possible for binary stars (mass; 3 stars in the sample)
%and bright stars (radius and $T_{\rm eff}$; 10 stars). 

 \end{abstract}

\firstsection % if your document starts with a section,

\section{Why are fundamental parameters important?}

Fundamental parameters are critical for the interpretation 
of both the exoplanet and asteroseismic data from CoRoT and {\em Kepler}. %  \cite{basu09}.
These space missions will provide a huge leap forward in our understanding 
of the interior physics of stars. This is possible by comparing 
the observed oscillation frequencies with theoretical pulsation models. 
It will allow us to examine how we can improve 
the approximations of the physics in the evolution models. % including mixing length theory and diffusion processes. 
To limit the range of models we need reliable estimates 
of the fundamental parameters of the target stars.
Also, characterization of stars hosting exoplanets is important to understand the properties of transiting systems.
Since the targets of CoRoT and especially {\em Kepler} are faint we must
use indirect methods. We will compare direct and indirect methods and
determine to what extent we can constrain $T_{\rm eff}$, mass and radius.
% The results will be published as Bruntt \etal\ (MNRAS, in prep.).

\begin{figure}[ht]
\begin{center}
\includegraphics[width=6.7cm]{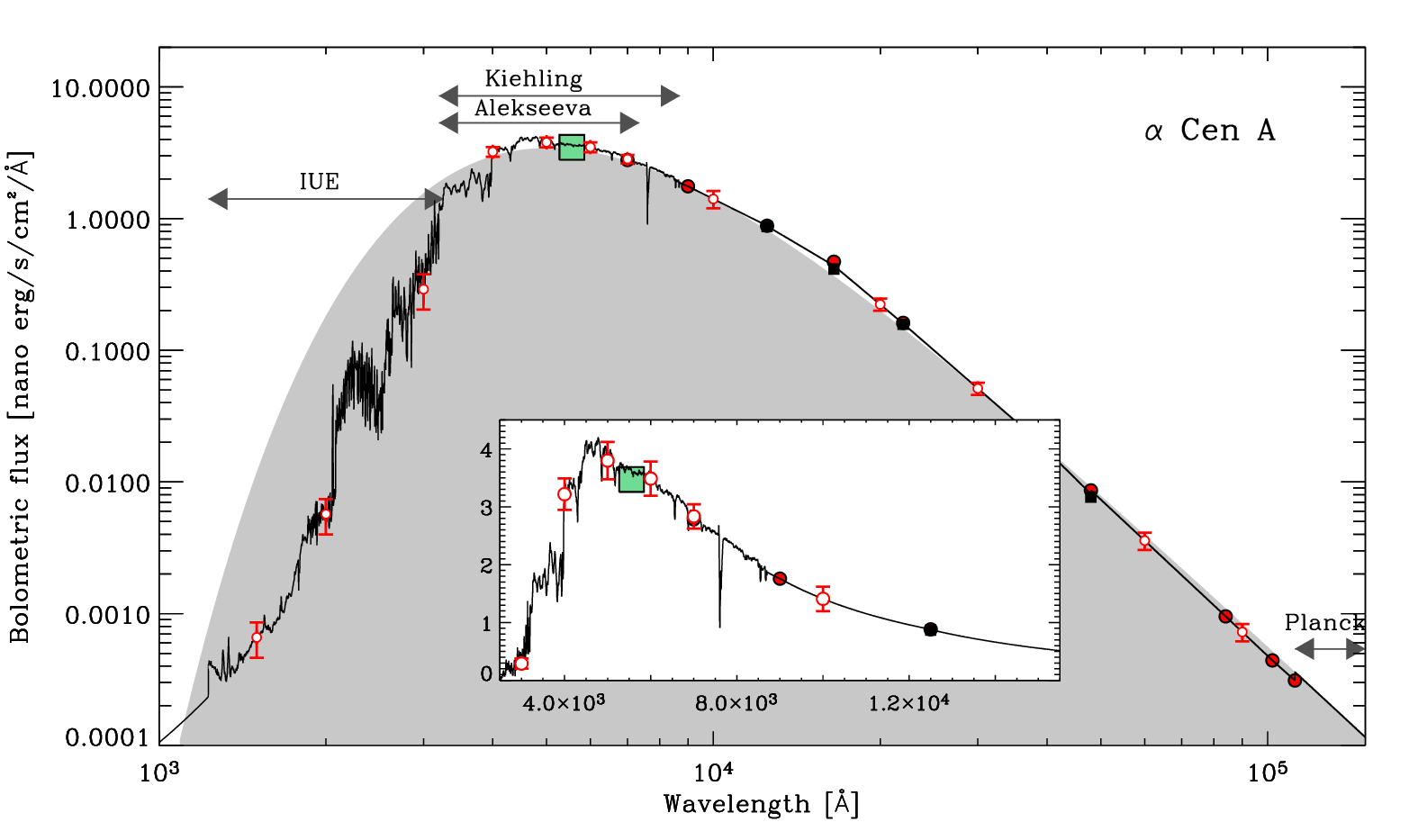} % updated 13 june 2009
%\hspace*{0.1cm}
\includegraphics[width=6.7cm]{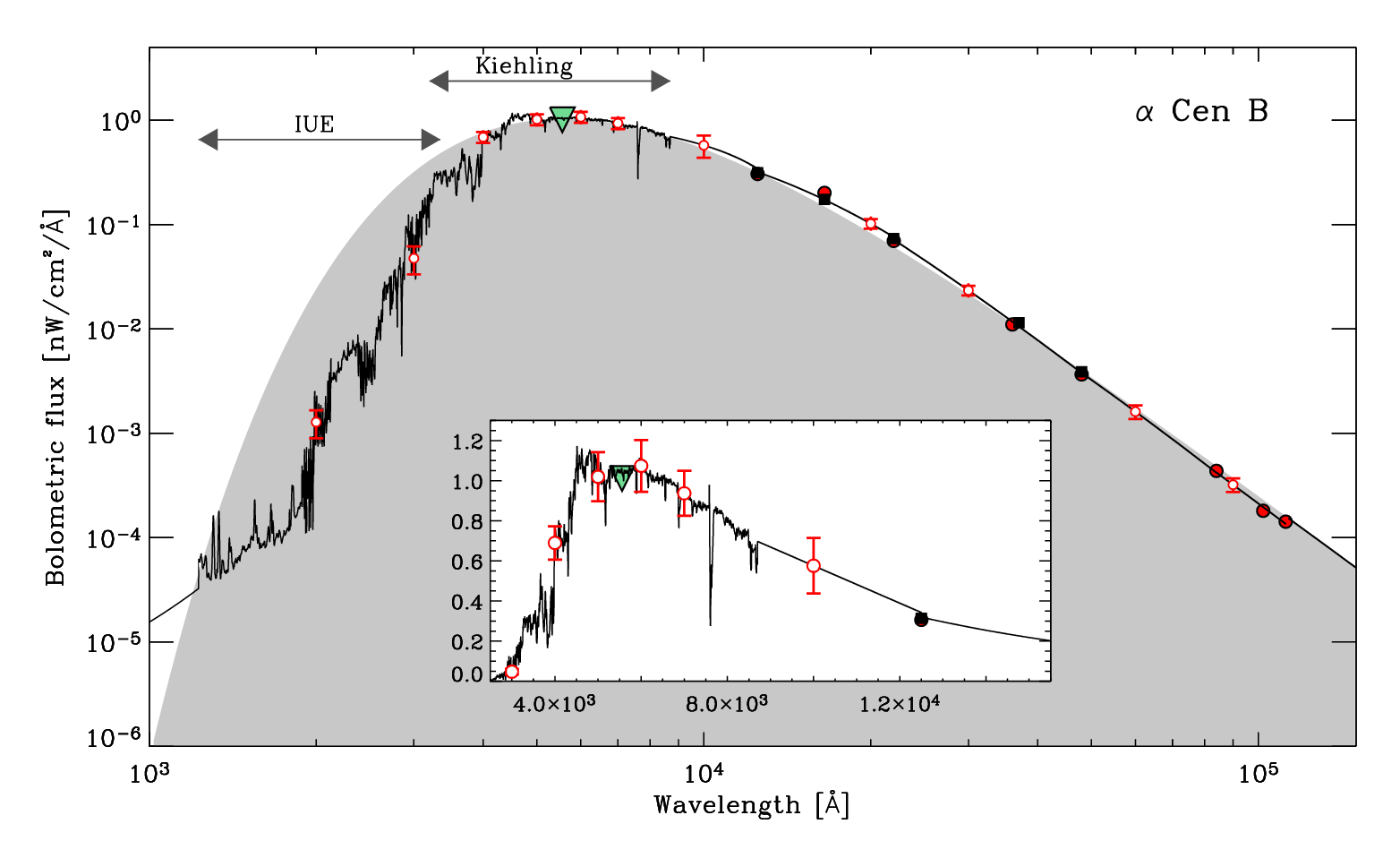} %% beta Vir
\caption{The data used to determine the bolometric flux of 
$\alpha$~Cen~A$+$B . IUE and ground-based spectrophotometric
data are shown with thin lines and filled symbols are broad-band fluxes. 
   \label{fig:flux}}
\end{center}
\end{figure}

\begin{figure}[ht]
\begin{center}
\includegraphics[width=6.6cm]{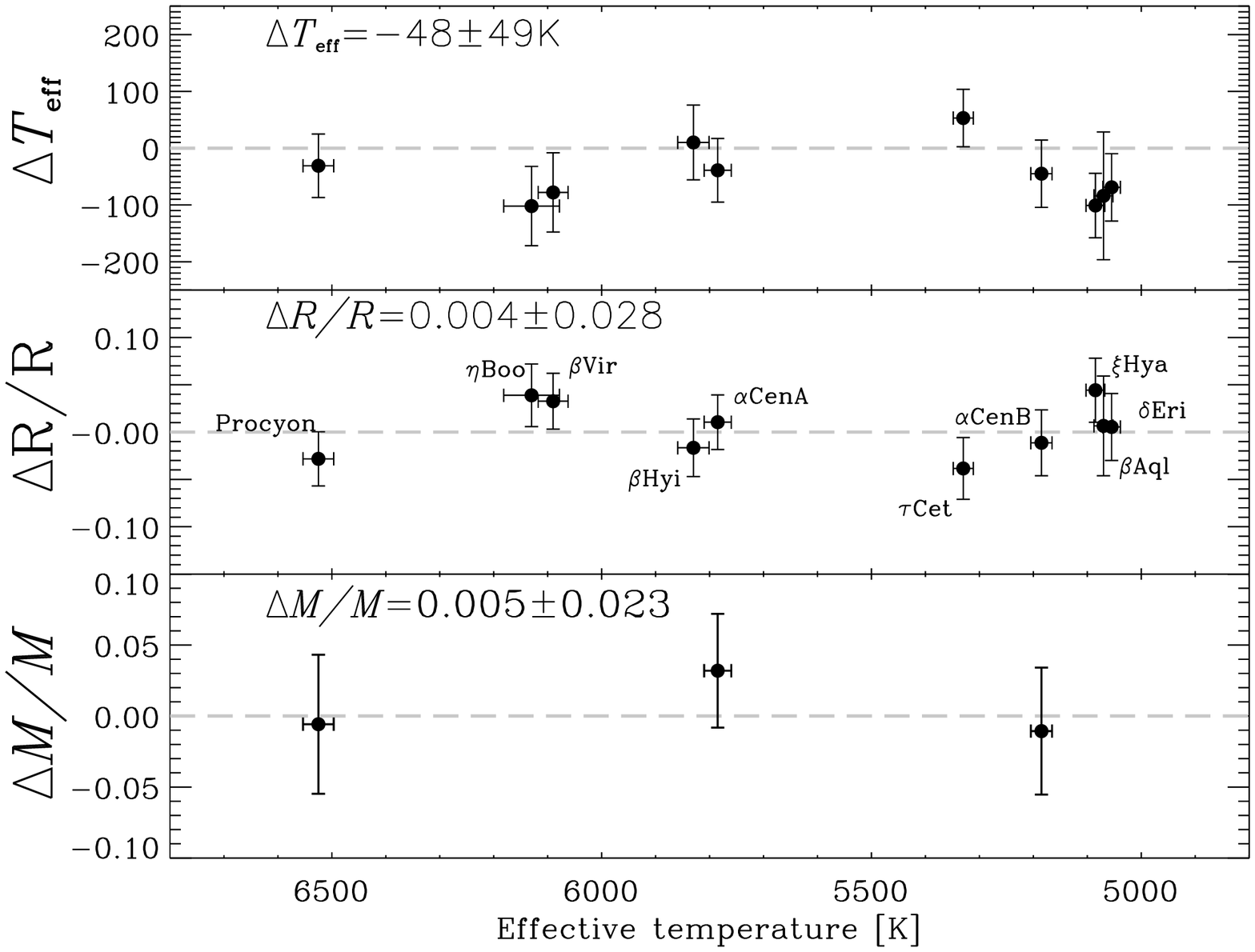}
\hspace*{0.1cm}
\includegraphics[width=6.6cm]{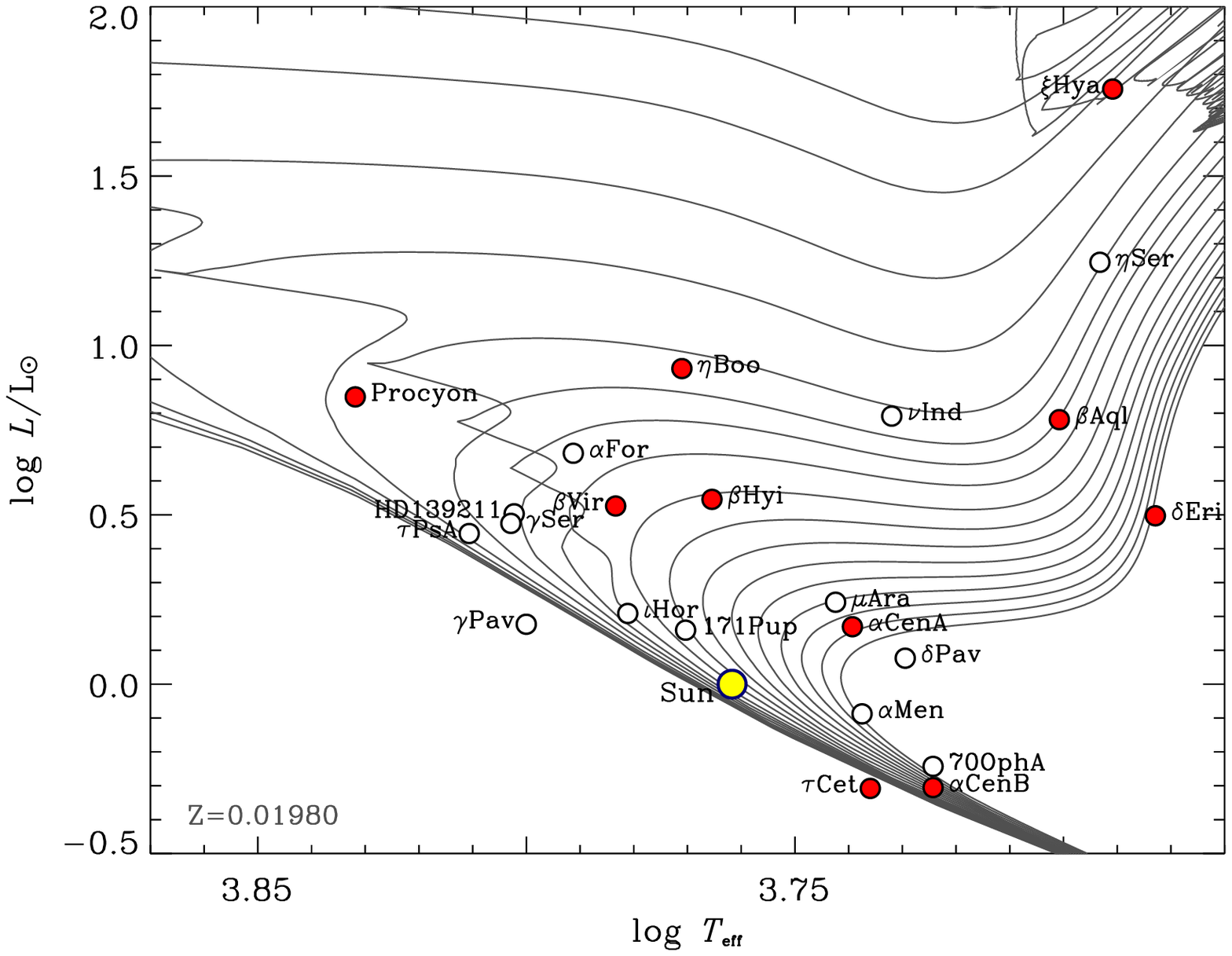}
\caption{
{\em Left panel:} Comparison of $T_{\rm eff}$, $R$ and $M$ from direct and indirect methods.
{\em Right panel:} H-R diagram with ASTEC tracks.
% Stars marked by solid circles have measured angular diameters. % interferometrically measured radii.
Solid circles are stars with measured angular diameters. % interferometrically measured radii.
   \label{fig:hr}}
\end{center}
\end{figure}

% -----------------------------------------
% -----------------------------------------
% NEW PAGE INSERTED MANUALLY
% -----------------------------------------
% -----------------------------------------
%\newpage
% -----------------------------------------
% -----------------------------------------
% NEW PAGE INSERTED MANUALLY
% -----------------------------------------
% -----------------------------------------

% \section
\vspace{0.3cm}
{\begin{center} 
{\underline {\bf {${\mathbf T_{\rm eff}}$ from spectroscopy with 50~K accuracy}}}
\end{center}

We compared two methods to determine $T_{\rm eff}$ of 10 bright solar-type stars. 
(1) We used measured angular diameters from the literature
combined with the bolometric flux (Fig.~\ref{fig:flux})
yielding $T_{\rm eff}$ from its basic definition. 
These results are nearly model-independent; only the limb-darkening is from models.
(2) We made a ``classical'' spectroscopic analysis of 100s
of Fe\,{\sc i} lines requiring that lines with a range of different
excitation potentials and line strengths yield the same abundance. 
We use the VWA tool (\cite{bruntt08, bruntt09}) 
employing 1D~LTE MARCS atmospheric models
(\cite{gustafsson08}; spectra are from HARPS@ESO except $\eta$~Boo observed with FIES@NOT). 
%% The analysis is done relative to a Solar spectrum (\cite{bruntt08})
As shown in Fig.\,\ref{fig:hr} 
(left top panel) the mean difference is $\Delta T_{\rm eff} = -48\pm49$\,K (rms scatter),
and there is no significant correlation with $T_{\rm eff}$. 
We thus claim that after correcting for the $\Delta T_{\rm eff}$ 
offset we can determine $T_{\rm eff}$ from a high-quality
spectrum to $\simeq50$~K in the spectral range from 
Procyon~A (F5) to $\alpha$~Cen~B (K1). 
The stars are shown in the H-R diagram in Fig.\,\ref{fig:hr}.

\vspace{0.3cm}
{\begin{center} 
{\underline {\bf {{Radius with 3\% accuracy without interferometry}}}}\\
\end{center}

We determined the radii of the stars using a direct and an indirect method:
(1) We combined the measured angular diameters from the literature with
the updated parallaxes from \cite{leeuwen07}. 
(2) We combined the spectroscopic $T_{\rm eff}$ with
the luminosity through $L/L_\odot = (R/R_\odot)^2 \, (T_{\rm eff}/T_{{\rm eff};\odot})^4$.
The luminosities were determined from the $V$ magnitude, BC from \cite{girardi02},
and the parallax from \cite{leeuwen07}. 
The comparison of the two methods to determine $R$ 
is shown in Fig.~\ref{fig:hr} in the left middle panel.
The agreement is good and shows that radii can be determined 
from indirect methods to $\simeq3$\% (1-$\sigma$ uncertainty).

\vspace{0.3cm}
{\begin{center} 
{\underline {\bf {{Mass with 4\% accuracy for single stars}}}}
\end{center}
}

Three of the stars are members of binary systems and have
well-determined masses ($<2$\%). 
For all stars the radius has been measured ($<2$\%) from interferometry 
and the mean density is inferred from asteroseismic data (large separation)
by scaling from the Sun. Combining the radius and density we get the
mass to $\simeq4$\%. The left lower panel in Fig.~\ref{fig:hr} compares the mass for
the two methods (Procyon~A and $\alpha$~Cen~A$+$B). Although the
number of stars is small it is reassuring that the agreement is
good ($2$\% rms scatter).

\end{document}